\documentclass[letter,oldversion,longauth]{aa} 
\usepackage{graphicx}
\usepackage{txfonts}
\def\ms{\hbox{\,m\,s$^{-1}$}}         
\def\m2s2{\hbox{\,m$^{2}$\,s$^{-2}$}} 
\def\kms{\hbox{\,km\,s$^{-1}$}}       
\def\vsini{\hbox{$v$\,sin\,$I$}}      
\def\sini{\hbox{sin\,$I$}}      

\begin{document}
\title{Transiting exoplanets from the CoRoT space mission}

\subtitle{III. The spectroscopic transit of CoRoT-Exo-2b with SOPHIE and HARPS
             \thanks{Observations made with SOPHIE spectrograph at Observatoire de 
	     Haute Provence, France (PNP.07A.MOUT) and HARPS spectrograph at ESO La Silla Observatory 
	     (079.C-0127(F)). The CoRoT space mission, launched on December 27th 2006, has been developed and is
	     operated by CNES, with the contribution of Austria, Belgium, Brasil, ESA, Germany, and Spain.}}

\author{
Bouchy, F. \inst{1}
\and Queloz, D. \inst{2}
\and Deleuil, M. \inst{3}
\and Loeillet, B. \inst{1,3}
\and Hatzes, A.P. \inst{4} 
\and Aigrain, S. \inst{5}
\and Alonso, R. \inst{3}
\and Auvergne, M. \inst{6} 
\and Baglin,~A. \inst{6} 
\and Barge, P. \inst{3}
\and Benz, W. \inst{7}
\and Bord\'e, P. \inst{8} 
\and Deeg, H.J. \inst{9}  
\and De la Reza, R. \inst{10}   
\and Dvorak, R. \inst{11} 
\and Erikson, A. \inst{12} 
\and Fridlund,~M. \inst{13} 
\and Gondoin, P. \inst{13} 
\and Guillot, T. \inst{14} 
\and H\'ebrard, G. \inst{1} 
\and Jorda, L. \inst{3}
\and Lammer, H. \inst{15} 
\and L\'eger, A. \inst{8}  
\and Llebaria, A. \inst{3} 
\and Magain,~P. \inst{16} 
\and Mayor, M. \inst{2}
\and Moutou, C. \inst{3}
\and Ollivier, M. \inst{8} 
\and P\"atzold, M. \inst{17} 
\and Pepe, F. \inst{2}
\and Pont, F. \inst{2}
\and Rauer, H. \inst{12,19}  
\and Rouan, D. \inst{6} 
\and Schneider, J. \inst{18} 
\and Triaud, A.H.M.J. \inst{2}
\and Udry, S. \inst{2}
\and Wuchterl, G. \inst{4} }

\offprints{\email{bouchy@iap.fr}}

\institute{
Institut d'Astrophysique de Paris, UMR7095 CNRS, Universit\'e Pierre \& Marie Curie, 98bis Bd Arago, 75014 Paris, France
\and
Observatoire de Gen\`eve, Universit\'e de Gen\`eve, 51 Ch. des Maillettes, 1290 Sauverny, Switzerland
\and
Laboratoire d'Astrophysique de Marseille, CNRS UMR 6110, Traverse du Siphon, 13376 Marseille, France
\and
Th{\"u}ringer Landessternwarte Tautenburg, Sternwarte 5, 07778 Tautenburg, Germany
\and
School of Physics, University of Exeter, Stocker Road, Exeter EX4 4QL, United Kingdom
\and
LESIA, CNRS UMR 8109, Observatoire de Paris, 5 place J. Janssen, 92195 Meudon Cedex, France
\and
Physikalisches Institut Universit\"at Bern, Sidlerstrasse 5, 3012 Bern, Switzerland
\and
Institut d'Astrophysique Spatiale, Universit\'e Paris XI, 91405 Orsay, France
\and
Instituto de Astrofisica de Canarias, E-38200 La Laguna, Tenerife, Spain
\and
Observat\'orio Nacional, Rio de Janeiro, RJ, Brazil
\and
Institute for Astronomy, University of Vienna, T\"urkenschanzstr. 17, A-1180 Vienna, Austria
\and
Institute of Planetary Research, DLR, Rutherfordstr. 2, 12489 Berlin, Germany
\and
Research and Scientific Support Department, European Space Agency, ESTEC, 2200 Noordwijk, The Netherlands 
\and
Observatoire de la C\^ote d'Azur, Laboratoire Cassiop\'ee, CNRS UMR 6202, BP 4229, 06304 Nice Cedex 4, France
\and
Space Research Institute, Austrian Academy of Sciences, Schmiedlstrasse 6, A-8042 Graz, Austria 
\and
Institut d'Astrophysique et de G\'eophysique, Universit\'e de Li\`ege, All\'ee du 6 Ao\^ut 17, Sart Tilman, Li\`ege 1, Belgium
\and
Institute for Geophysics and Meteorology, K\"oln University, Albertus-Magnus-Platz, 50923 Cologne, Germany
\and
LUTH, Observatoire de Paris-Meudon, 5 place J. Janssen, 92195 Meudon Cedex, France
\and
Center for Astronomy and Astrophysics, TU Berlin, Hardenbergstr. 36, 10623 Berlin
}

\date{Received ; accepted }

 
\abstract
{We report on the spectroscopic transit of the massive hot-Jupiter CoRoT-Exo-2b observed with 
the high-precision spectrographs SOPHIE and HARPS. By modeling the radial velocity anomaly 
occurring during the transit due to the Rossiter-McLaughlin (RM) effect, we determine the 
sky-projected angle between the stellar spin and the planetary orbital axis to be close to 
zero $\lambda=7.2\pm4.5$ deg, and we secure the planetary nature of CoRoT-Exo-2b. 
We discuss the influence of the stellar activity on the RM modeling. Spectral analysis 
of the parent star from HARPS spectra are presented.}


\titlerunning{The spectroscopic transit of CoRoT-Exo-2b}

\authorrunning{F. Bouchy et al.}

\maketitle

%

\section{Introduction}

Measurement of the spectroscopic signal during the transit of an exoplanet in front of its host star
-- known as the Rossiter-McLaughlin (RM) effect -- provides an assessment the trajectory of the planet 
across the stellar disk and, more precisely, the sky-projected angle between the planetary orbital 
axis and the stellar rotation axis. This misalignment angle, denoted by $\lambda$, is a fundamental 
property of planetary systems that provides clues about the process of planet migration. 
Among the 30 transiting exoplanets known so far, $\lambda$ has been reported for only 5 exoplanets 
(HD209458b, Queloz et al. \cite{queloz}; HD189733b, Winn et al. \cite{winn06}; HAT-P-2, Winn et al. 
\cite{winn07}, Loeillet et al. \cite{loeillet}; HD149026b, Wolf et al. \cite{wolf}, and TrES-1, 
Narita et al. \cite{narita}). For all of these cases, $\lambda$ is close to zero, as in the 
solar system, and the stellar rotation is prograde relative to the planet orbit. 
Such measurements should be extended to other transiting systems to understand whether this 
degree of alignment is typical.   

The massive hot-Jupiter CoRoT-Exo-2b (Alonso et al. \cite{alonso}) was revealed as planetary 
candidate by the CoRoT space mission (Baglin et al. \cite{baglin}) and its planetary nature 
and mass was established thanks to ground-based facilities, including high-precision spectrographs 
SOPHIE (Bouchy et al. \cite{bouchy}) and HARPS (Mayor et al. \cite{mayor}). 
This second CoRoT exoplanet is a 3.3 Jupiter-mass planet orbiting an active G7 dwarf star (mv=12.6) 
every 1.743 days. 
We report here the measurements of the spectroscopic transit observed with both SOPHIE and 
HARPS spectrographs. These observations were made simultaneously with the space-based 
photometry with CoRoT. Such simultaneous monitoring is useful to assess anomalies in the 
transit parameters due to star spots or transient events.\\

Our data permits us to determine the sky-projected angle between the stellar spin and 
the planetary orbital axis, and it provids additional constraints on the orbital 
and physical parameters of the system. Furthermore, our data confirms and secures 
the planetary nature of the transiting body, excluding blending of an eclipsing binary 
with a third star as the cause of the observed shallow transits. We used HARPS 
spectra to perform the spectroscopic analysis of the parent star. 

\section{Observations}
We performed high-precision radial velocity observations of CoRoT-Exo-2 (mv=12.6) 
with the SOPHIE spectrograph, based on the 1.93-m OHP telescope (France), and 
the HARPS spectrograph, based on the 3.6-m ESO telescope (Chile). 
These two instruments are cross-dispersed, fiber-fed, echelle spectrographs dedicated to 
high-precision Doppler measurements based on the radial velocity techniques of 
simultaneous-Thorium calibration.  
SOPHIE was used with its high efficiency mode (spectral resolution R=40,000). 
We reduced HARPS and SOPHIE data with the same pipeline based on the 
cross-correlation techniques (Baranne et al. \cite{baranne}; Pepe et al. \cite{pepe}). 
We observed CoRoT-Exo-2 with SOPHIE on 16 July 2007 and 
with HARPS on 1 September 2007. The exposure times were respectively 10 and 20 
minutes on HARPS and SOPHIE corresponding to S/N per pixel at 550 nm of  
16 and 25, respectively. We obtained the radial velocities by weighted cross-correlation 
with a numerical G2 mask constructed from the Sun spectrum atlas including up to 3645 lines. 
We eliminated the first 8 blue spectral orders containing only noise. 
Radial velocities are given in Table~\ref{tab:rv} and displayed in Fig.~\ref{fig:rv}. 

\begin{table}
\centering                          
\caption{Radial velocity measurements of CoRoT-Exo-2 obtained by HARPS and SOPHIE during 
the transit. BJD is the Barycentric Julian Date.}  
\begin{tabular}{lll}
\hline\hline
BJD&  RV & Uncertainty \\
-2400000   & [km\,s$^{-1}$] & [km\,s$^{-1}$]   \\
\hline
\multicolumn{3}{c}{SOPHIE 2007-07-16}\\
\hline
54298.4641 &   23.341 & 0.026 \\
54298.4862 &   23.285 & 0.027 \\
54298.5030 &   23.369 & 0.028 \\
54298.5198 &   23.378 & 0.027 \\
54298.5381 &   23.123 & 0.027 \\
54298.5550 &   22.926 & 0.028 \\
54298.5714 &   22.891 & 0.029 \\
54298.5879 &   23.023 & 0.030 \\
\hline 
\multicolumn{3}{c}{HARPS 2007-09-01} \\
\hline
54345.5225 & 23.371   &   0.020 \\
54345.5298 & 23.371   &   0.019 \\
54345.5371 & 23.392   &   0.018 \\
54345.5444 & 23.360   &   0.018 \\
54345.5517 & 23.347   &   0.019 \\
54345.5590 & 23.370   &   0.018 \\
54345.5663 & 23.456   &   0.018 \\
54345.5736 & 23.488   &   0.020 \\
54345.5809 & 23.496   &   0.022 \\
54345.5883 & 23.394   &   0.019 \\
54345.5956 & 23.271   &   0.018 \\
54345.6029 & 23.190   &   0.018 \\
54345.6124 & 23.060   &   0.018 \\
54345.6197 & 23.005   &   0.017 \\
54345.6270 & 22.939   &   0.018 \\
54345.6343 & 22.928   &   0.018 \\
54345.6417 & 23.034   &   0.019 \\
54345.6490 & 23.093   &   0.019 \\
54345.6563 & 23.107   &   0.019 \\
54345.6636 & 23.090   &   0.019 \\
54345.6709 & 23.034   &   0.020 \\
\hline
\end{tabular}
\label{tab:rv}
\end{table}

\section{Rossiter-McLaughlin modeling}

\begin{figure*}
\centering
\includegraphics[width=12cm]{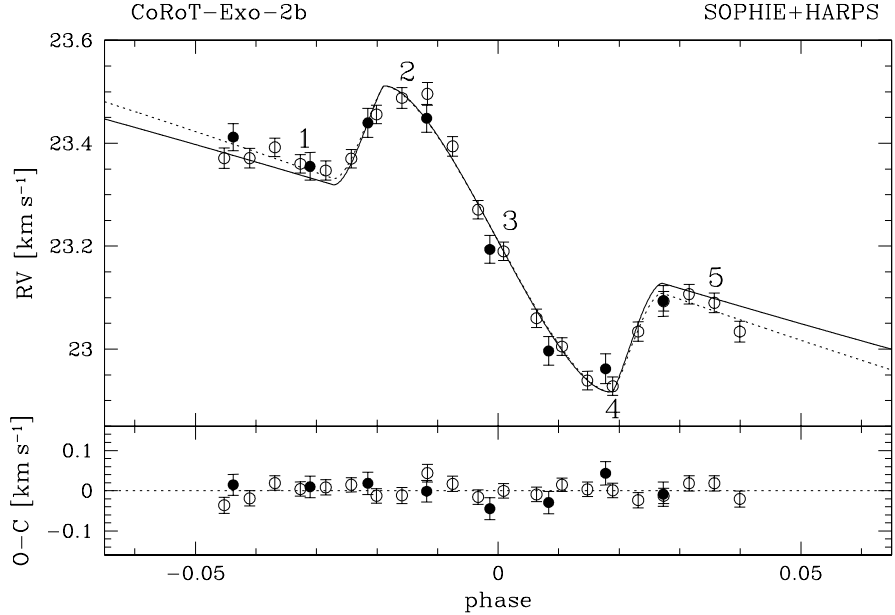}
\caption{Phase-folded radial velocity measurements of CoRoT-Exo-2 during the transit of the planet 
with SOPHIE (dark circle) and HARPS (open circle). The solid line corresponds to the Rossiter-McLaughlin 
model ajusted to these data assuming the semi-amplitude $K$=563 {\ms} from Alonso et al. (\cite{alonso}). 
The dotted line corresponds to the Rossiter-McLaughlin 
model with $K$ as free parameters.}  
\label{fig:rv}
\end{figure*}

The RM effect corresponds to a distortion of the spectral lines 
observed during a planetary transit due to stellar rotation. The transiting body hides some of the 
velocity components that usually contribute to line broadening resulting in an 
Doppler-shift anomaly (see Otha et al. \cite{otha}; Gim\'enez et al. \cite{gimenez06b}; Gaudi \& Winn 
\cite{gaudi}). 

To model this RM effect, we used the analytical approach 
developed by Otha et al. (\cite{otha}). The complete model has 12 parameters: the orbital period $P$; 
the mid-transit time $T_c$; the eccentricity $e$; the angle between the node and periastron $\omega$; 
the RV semi-amplitude $K$; the velocity zero point $V_0$ (these first six are the standard 
orbital parameters); the radius ratio $r_p/R_s$; the orbital semi-major axis to stellar radius 
$a/R_s$ (constrained by the transit duration); the sky-projected angle between the stellar spin axis 
and the planetary orbital
axis $\lambda$; the sky-projected stellar rotational velocity \vsini; the orbital inclination 
$i$; and the stellar limb-darkening coefficient $\epsilon$. For our purpose, we started with the 
orbital parameters and photometric transit parameters as derived by Alonso et al. (\cite{alonso}). 
We fixed the linear limb-darkening coefficient $\epsilon$=0.78, based on Claret (\cite{claret}) 
tables for filter $g'$ and for the stellar parameters derived in Sect.~\ref{stellar}.   
Our free parameters are then $\lambda$ and \vsini. We introduced two additional 
parameters: the offset velocity of HARPS and SOPHIE, $\Delta_{HARPS}$ and $\Delta_{SOPHIE}$, 
which differ from $V_0$ due to the stellar activity. 
We determined the {\vsini} independently from SOPHIE cross-correlation functions (CCFs) to 
be $9.5\pm1.0$ {\kms} and from HARPS CCFs to be $10.7 \pm0.5$ {\kms} with the calibration 
techniques described by Santos et al. (2002). However, we decided to leave it as free 
parameter in our fit. 

The result of our fit, displayed in Fig.~\ref{fig:rv} and listed in Table~\ref{tab:rm}, first 
shows that the stellar rotation is prograde relative to the planet orbit. 
During the first part of the transit the starlight is redshifted, indicating that the planet is 
in front of the approaching (blueshifted) half of the stellar disk. During the second part of 
transit, the sign is reversed as the planet moves to the receding (redshifted) half of the stellar disk.  
The sky-projected angle between the stellar spin axis and the planetary orbital
axis $\lambda$ is close to zero. The projected rotation velocity of the star 
$v$\,sin\,$I$ determined by our RM fit (11.85$\pm$0.5 \kms) seems slightly larger than our spectroscopic 
determination (2-$\sigma$ greater). Previous studies by Winn et al. (\cite{winn05}) showed that 
the $v$\,sin\,$I$ measured with Otha formulae was biased toward larger values by 
approximatively 10\%. But, as already suggested by Loeillet et al. (\cite{loeillet}), it may be due 
to the differential rotation of the star from equator to pole.  
Considering the exoplanet crosses the star near its 
equatorial plan, the fitted $v$\,sin\,$I$ corresponds to the maximum value. 
Note that if we fix $v$\,sin\,$I$ at the spectroscopic value, it does 
not change the value of the fitted $\lambda$ angle.

We made the 2 epochs of RM observations at a minimum stellar flux (see Fig.~1 of Alonso et al. 
\cite{alonso}), indicating that the stellar spots were at their maximum phase of visibility. 
Following the Saar \& Donahue (\cite{saar}) relation giving the expected RV jitter as 
a function of {\vsini} and spot filling factor, we found that CoRoT-Exo-2 is expected to present 
RV variations of up to 200 {\ms} peak-to-peak with a period of 4.5 days. 
The standard deviation of RV residuals (56 {\ms}) found by Alonso et al. (\cite{alonso}) is in agreement  
with this value. Such an activity-related RV variation should then change locally the apparent slope 
in the RV orbital curve. The maximum effect occurs at the maximum phase of stellar spot visibility, 
and should induce an apparent increase in the semi-amplitude $K$ of up to 40 {\ms}.    
This explains why our fit in Fig.~\ref{fig:rv} is not perfect outside of the transit. 
If we increase $K$ in our fit or leave it as a free parameter, it significantly improves the fit 
and slightly decreases the value of $v$\,sin\,$I$ and $\lambda$ (see Table~\ref{tab:rm}). 
            
We also did a combined fitting of the photometry and the whole set of RV measurements. 
On each of the out-of-transit measurements, we inserted an additional error on the RV data 
to take the stellar activity into account. We chose this value as 56 m\,s$^{-1}$, 
corresponding to the standard deviation found by Alonso et al (\cite{alonso}). 
This correction is justified since the action of activity 
on the points taken at random out-of-transit phases can be assumed as random 
for these points, while during transit we have sets of points with the same 
activity level throughout. The fitting was done using a Markov Chain Monte Carlo (MCMC) 
with a Metropolis-Hastings Algorithm for the decision process. 
We used the models of Gim\'enez \cite{gimenez06a} and \cite{gimenez06b} for photometry and 
spectroscopic transits respectively. A quadratic law of limb darkening 
was used. For the photometry, we used the fitted parameters found by Alonso et al. (\cite{alonso}). 
For the spectroscopy parameters, we chose the $V$-band, from tables published by Claret 
2000 for the stellar parameters derived in Sect.~\ref{stellar} ($u_+=0.748$, $u_-=0.256$).
The MCMC was performed over 20\,000 accepted steps after 5\,000 steps of a burn-in period. 
The result of the combined fit is presented in Table~\ref{tab:rm}, and is in full agreement with 
the other approachs.

\begin{table}
\centering                          
\caption{System parameters of CoRoT-Exo-2. The reduced $\chi^2$ was computed 
assuming 24 degrees of freedom.}  
\begin{tabular}{ll}
\hline\hline
\multicolumn{2}{c}{Fixed parameters from Alonso et al. \cite{alonso}} \\
\hline
$P$ & 1.7429964 days  \\
$T_c$ & 54237.53562  \\
$e$  & 0.0 \\
$V_0$ & 23.245 km\,s$^{-1}$  \\ 
$r_p/R_s$  &  0.1667  \\
$a/R_s$  &  6.70  \\
$i$ &  87.84 deg \\
$\epsilon$ & 0.78 (from Claret) \\
\hline
\multicolumn{2}{c}{Adjusted parameters with $K$=563 m\,s$^{-1}$} \\
\hline
$v$\,sin\,$I$ & 11.85$\pm$ 0.50 km\,s$^{-1}$ \\
$\lambda$ & 7.2$\pm$4.5 deg \\
$\Delta_{HARPS}$  & -21.5 $\pm$ 5 m\,s$^{-1}$ \\
$\Delta_{SOPHIE}$  & +21.5 $\pm$ 12 m\,s$^{-1}$\\
reduced $\chi^2$ & 1.43 \\
\hline
\multicolumn{2}{c}{Adjusted parameters with $K$ as free parameter} \\
\hline
$K$ & 656$\pm$27 m\,s$^{-1}$ \\
$v$\,sin\,$I$ & 11.25$\pm$ 0.45 km\,s$^{-1}$ \\
$\lambda$ & 5.0$\pm$4.0 deg \\
$\Delta_{HARPS}$  & -25.0 $\pm$ 4.5 m\,s$^{-1}$ \\
$\Delta_{SOPHIE}$  & +25.5 $\pm$ 11 m\,s$^{-1}$\\
reduced $\chi^2$ & 1.01 \\
\hline
\multicolumn{2}{c}{Combined MCMC fit} \\
\hline
$K$ & 613$\pm$14 m\,s$^{-1}$ \\
$v$\,sin\,$I$ & $11.46 ^{+ 0.29}_{-0.44}$ km\,s$^{-1}$ \\
$\lambda$ & 7.1$\pm$5.0 deg \\
$\Delta_{HARPS}$  & -22.5 $\pm$ 4.5 m\,s$^{-1}$ \\
$\Delta_{SOPHIE}$  & +23.5 $\pm$ 11 m\,s$^{-1}$\\
reduced $\chi^2$ & 1.10 \\
\hline
\end{tabular}
\label{tab:rm}
\end{table}

\begin{figure}
\centering
\includegraphics[width=8.5cm]{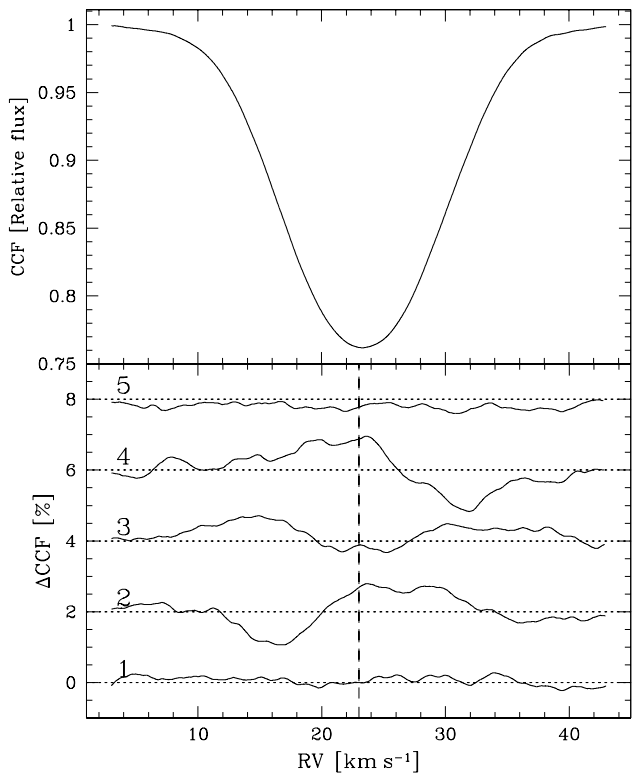}
\caption{(Top) Averaged cross-correlation function of CoRoT-Exo-2. (Bottom) 
Cross-correlation differences computed at 5 different epochs (see text) illustrating
the behavior of the spectral lines during the transit.} 
\label{fig:diffccf}
\end{figure}

The cross-correlation function (CCF) corresponds more or less to an average of all 
the spectral lines (see top of Fig.~\ref{fig:diffccf}). In order to characterize the 
behavior of the spectral lines  
during the transit, we computed the difference between the HARPS CCFs corrected from the 
orbital velocity and a reference CCF taken out of the transit (more exactly an average 
of the 3 first exposures). 
This difference was computed at 5 epochs identified and labeled in Fig.~\ref{fig:rv} :  
(1) just before the ingress, (2) maximum of the RM effect, (3) mid-transit epoch, (4) 
minimum of the RM effect, (5) just after egress. 
These differences $\Delta CCF=CCF_{\rm REF}-CCF_\#$ are displayed in Fig.~\ref{fig:diffccf} and clearly 
show the spectroscopic anomaly shifting from the blue side (2) to the red side (4) of the 
CCF. During the transit, the depth or contrast of the CCFs is systematically larger, 
reflecting the renormalization effect of the CCF, which maintains a constant surface. 

The observation of the spectroscopic transit of CoRoT-Exo-2b allows us to confirm definitively 
that the transiting candidate provided by CoRoT occurred at the central star 
(and not at a background star inside the CoRoT PSF).
Furthermore, if we assume that the system is not diluted by an other star inside the HARPS or SOPHIE 
PSF, the RM anomaly reveals that the transiting body has a planetary size (from the RM anomaly amplitude) 
and planetary mass (from the RV slope outside the transit). In the case of an eclipsing binary whose light 
is diluted with a brighter third star, one should assume that the spectral lines of the fainter 
eclipsing binary move relative to the lines of the bright star and thus change the blended line-profiles. 
In such a configuration, one should consider not only the flux ratio but the {\vsini}, velocity zero 
point, and spectral type of the two systems. In our present case, we did not find a configuration 
of a blended eclipsing binary that could simultaneously reproduce the RV anomaly and 
the photometric light curve. Furthermore, we computed RVs using different cross-correlation mask without 
significant changes in the shape and amplitude of the RM anomaly. We, thus, consider that the spectroscopic 
transit confirms and secures the planetary nature of the transiting body.

\section{Spectroscopic analysis of CoRoT-Exo-2}
\label{stellar}

We performed the spectroscopic analysis of the parent star using the HARPS spectra. 
We corrected individual spectra from the stellar velocity, rebinned to a constant wavelength step 
of 0.02 {\AA}, and co-added spectral order per spectral order giving a S/N per pixel at 550 nm 
of about 80. We determined the effective temperature first from the analysis of the H$\alpha$ line 
wings, providing a temperature of 5450$\pm$120~K. 
In spite of the quite low S/N of the combined spectra, it appears that the star is at the border of the temperature 
domain in which the H$\alpha$ line wings are a good temperature indicator (from 5500 to 8500 K). 
We then checked this result with other methods. We performed synthetic spectra fitting using LTE MARCS atmosphere 
models (Gustafsson et al. \cite{gusta05}), 
which are well adapted for this range of temperature. We compared the synthetic spectra, previously convolved by the 
instrumental profile and a rotational profile with the $v$\,sin\,$I$ value previously measured, 
to the observed one. The best-fit model yields a slightly higher temperature, but is still in agreement 
with the H$\alpha$ estimate. Another analysis, using equivalent width measures of FeI and FeII lines, 
was also carried out and yields similar results. The adopted stellar parameters are T$_{\rm eff}$=5625$\pm$120~K, 
log$g$=4.3$\pm$0.2 and [M/H]=0.0$\pm$0.1, which correspond to a G7V type star with a solar metallicity. 
With these values, we derived the star's luminosity and mass with {\sl StarEvolv} stellar evolution models 
(Siess \cite{siess}; Palacios, private communication). We combined these estimates of the star's mass to the 
$M_s^{1/3}/R_s$ value provided by the light curve analysis to derive the final star's mass and radius 
values in a consistent way between spectroscopic and photometric analyses. The method allows us to get rid 
of the large uncertainty that affects the estimate of the 
gravity and to take advantage of the excellent quality of the light curve. 
The method will be detailed in a forthcoming paper devoted to the fundamental parameters of the first CoRoT 
planet host stars, based on {\sl UVES} spectra. 
The adopted stellar mass is 0.97$\pm$0.06 M$_\odot$ and the stellar radius is 0.90$\pm$0.02 R$_\odot$.
Interestingly, the solar-like metallicity of the parent star and large radius of the planet
is consistent with the trend that heavy element content in the planet and stellar
metallicity are correlated (Guillot et al. \cite{guillot}).
According to stellar evolution models (Lebreton, private communication), the age of the star 
could be between 0.2 and 4 Gyr if the star is on the main sequence. However, the presence 
of the Li I absorption line and the strong emission line core in the CaII H and K lines, suggest 
that the star is still close the ZAMS and could be thus younger than 0.5 Gyr in full agreement 
with the observed stellar activity and the measured rotation period.

The knowledge of the main rotational period of CoRoT-Exo-2 determined from the light curve (4.54 days) 
and the spectroscopic $v$\,sin\,$I$ determined from HARPS and SOPHIE CCFs (10.3 km\,s$^{-1}$) 
may be used to independently estimate the minimum radius of R$_s${\sini}=
0.92 $R_\odot$ in very good agreement
with our previous determination based on spectral classification. We note that this estimate, based 
on the well-determined stellar rotation thanks to the high-precision CoRoT light curve, does not depend on any 
spectral classification. On the other hand, if we assume the stellar radius from spectral analysis, 
we can deduce that {\sini} is close to 1, indicating a further constraint on the alignment 
of orbit and stellar spin.

\section{Conclusions}

In addition to the previous 5 transiting exoplanets where $\lambda$ angle have been reported, 
CoRoT-Exo-2b presents a prograde orbit relative to the stellar rotation and an 
angle $\lambda=7.2\pm4.5$ deg, close to zero.  
Our observations illustrate and demonstrate the capability of extending the reach 
of the RM technique to relatively-faint host stars (mv$\ge$12) like the CoRoT targets 
even with a 2-m class telescope.

\begin{acknowledgements}
The authors wish to thank Xavier Bonfils and Gaspare Locurto for their 
precious help and support on HARPS observations.
We are grateful for the OHP staff support at the 1.93-m telescope. 
The German CoRoT team (TLS and Univ. Cologne) acknowledges the 
support of DLR grants 50OW0603 and 50OW0204. 
FB acknowledges the support of PLS230371. HJD acknowledges support by grants 
ESP2004-03855-C03-03 and ESP2007-65480-65480-C02-02 of the Spanish Education and 
Science ministry. The authors thank the referee for insightful comments and suggestions.\\
\end{acknowledgements}

\end{document}